# Cu$_2$O porous nanostructured films fabricated by positive bias sputtering deposition

Yiqi Zhu[1†], Ji Ma[2‡], Lei Zhou[2], Yang Liu[2], Meiping Jiang[1,2], Xianfang Zhu[3], Jiangbin Su[1,2,3*]

1 College of Electrical & Communication Engineering, Jiangsu University of Technology, Changzhou 213001, China

2 School of Mathematics & Physics, Changzhou University, Changzhou 213164, China

3 China-Australia Joint Laboratory for Functional Nanomaterials, Xiamen University, Xiamen 361005, China

†,‡ Both authors contributed equally to this work and should be considered as co-first authors.

* Corresponding author. Email: jbsu@cczu.edu.cn (J. B. Su)

## Abstract

In this work, the authors fabricated Cu$_2$O porous nanostructured films (PNFs) on glass slide substrates by the newly developed positive bias deposition approach in a balanced magnetron sputtering (MS) system. It was found that the surface morphology, crystal structure and optical property of the as-deposited products were greatly dependent on the applied positive substrate bias. In particular, when the substrate was biased at +50 V and +150 V, both of the as-prepared Cu$_2$O PNFs exhibited a unique triangular pyramids-like structure with obvious edges and corners and little gluing, a preferred orientation of (111) and a blue shift of energy band gap at 2.35 eV. Quantitative calculation results indicated that the traditional bombardment effects of electrons and sputtering argon ions were both negligible during the bias deposition in the balanced MS system. Instead, a new model of tip charging effect was further proposed to account for the controllable formation of PNFs by the balanced bias sputtering deposition.

# 1. Introduction

As one of the most common two kinds of copper oxides, cuprous oxide ($Cu_2O$) is an important p-type transition metal oxide semiconductor material. Due to the advantages of low-cost, non-toxicity and abundant copper sources and the potential applications in the fields of gas sensors [1,2], solar cells [3,4] and photocatalysts [5-7], $Cu_2O$ thin films have attracted great interest of researchers. In order to enhance the performances of the above $Cu_2O$-based surface-sensitive devices and materials, the researchers tend to prepare $Cu_2O$ thin films of porous or nanoporous structures [2,6]. In literature, people often concentrates on the porosity and surface area of porous or nanoporous films while the unique size or curvature effect [8,9] of nanoscale building units or nanostructures has not attracted sufficient attention. Based on this consideration, a special kind of porous thin films which consists of solid and/or hollow nanostructures, such as nanoparticles, nanoligaments, nanowires, nanoplates, nanocavities, nanopores or nanochannels, is proposed herein. We call such porous nanostructure-films or nanostructured films hereafter PNFs for short. Due to the unique size or curvature effect of building nanostructures along with the increased porosity and surface area, it is expected that such thin films of PNF structures may have some different or improved performances in contrast to the traditional porous and solid thin films [10,11]. Thus, it makes great sense to fabricate $Cu_2O$ PNFs with tunable building nanostructures and further study their corresponding properties.

In the existing literature, the main preparation methods of $Cu_2O$ films include direct oxidation [12-14], electrochemical deposition [15-17] and magnetron sputtering (MS) [18-21]. However, these methods have not been found for the preparation of $Cu_2O$ porous films, especially the $Cu_2O$ PNFs with excellent properties of nanoscale building units. Instead, people prefer to utilize the templating [22], sol-gel [23] and dealloying [24] methods to prepare porous films. However, all of these conventional methods for the preparation of porous films have disadvantages more or less. For example, the pore shape and size distribution of porous films prepared by templating is determined by the template and thus seem not flexible or less-diversity. The porous films prepared by sol-gel method have the faults of high shrinkage rate, poor adhesion, easy cracking and residual hydroxyl and carbon in the film. Dealloying was initially proposed for the fabrication of porous metal thick films and then was further developed to fabricate metal oxide PNFs by Su *et al* [10,11]. Nevertheless, both of them may introduce impurities to the porous films from the solution. Therefore, it is very imperative to find or develop a new approach for the preparation of PNFs to overcome the above shortcomings. As we all known, MS is a commonly used method to prepare high quality thin films at present. It will not introduce other impurities and has the advantages of simple equipment, easy control, large coating area and strong adhesion. In fact, porous metal thin films such as Ti [25] and porous semiconductor (composite) thin films such as Zn/ZnO [26] and $TiO_2$ [27] have been successfully prepared by direct-current MS method. Similarly, can we also use MS equipment or make some targeted improvement to realize the flexible preparation of $Cu_2O$ PNFs?

With the above considerations, in this work we successfully prepared various $Cu_2O$ PNFs by our newly developed positive bias sputtering deposition approach *via* adjusting the applied substrate bias voltage in a balanced MS system. It was observed that the surface morphology, crystal structure (or texture) and optical property of the as-prepared $Cu_2O$ PNFs were greatly dependent on the applied substrate bias voltage. A new mechanism of tip charging effect was proposed for a full explanation of the substrate bias voltage-dependent formation of PNFs while the traditional ion bombardment effect was demonstrated to be negligible.

## 2. Experimental section

All samples were deposited onto glass slides in a JGP500A mode radio-frequency balanced MS system. A Cu target of 99.99 wt.% purity was fixed to a 6 kGs magnetic cathode, which was connected with a 13.56 MHz RF power supply. The distance between the Cu target and the glass slide substrate was set to be ~15 cm. The chamber was evacuated to a base pressure of $5.0 \times 10^{-4}$ Pa and subsequently back-filled with a steady flow of high pure Ar (15 sccm, 99.999 wt.%) to maintain a chamber pressure of 0.10 Pa. Note that no external oxygen gas flow was introduced to the chamber due to the sufficient residual oxygen in the pre-pumped high vacuum chamber [28]. Before each deposition, a pre-sputtering of 10 min was regularly performed to remove the possible oxide layer on the target surface. Then formal sputtering deposition was carried out at a radio-frequency power of 80 W, which led to a deposition rate of ~0.045 nm/s. During the deposition, the substrate mounted on a 3-inch Cu substrate holder was biased using a direct-current power supply at 0 V, +50 V, +100 V, +150 V and +200 V, respectively. The film thicknesses were monitored by a quartz crystal oscillator and all controlled to be ~240 nm, which were further confirmed by a surface profile measuring equipment (Veeco Dektak 150). The as-achieved deposits were characterized by field-emission scanning electron microscope (FESEM,

ZEISS SUPRA 55), powder X-ray diffractometer (XRD, RIGAKU D/Max 2500 PC) and ultraviolet-visible (UV-vis) spectrophotometer (SHIMADZU UV-2450).

## 3. Results and discussion

### 3.1 Surface morphology

FESEM images in Fig.1 show the typical surface morphology of the $Cu_2O$ films prepared under different substrate bias voltage. It can be seen from the micrographs that all the $Cu_2O$ films have the characteristics of large area, well-distributed, smooth surface and PNF structure. However, the details of their building nanostructures are quite different. When no external bias voltage is applied ($V_s$=0 V), as shown in Fig. 1(a), the film presents a bicontinuous "ligament-channel" PNF structure. Most of the ligaments are curved ropes shape, short in length, glued to each other and only partly with obvious edges and corners. The average diameters of the ligaments and channels are ~27 nm and ~15 nm respectively. When the substrate is biased at +50 V, as shown in Fig. 1(b), nanoscale triangular pyramids with cut sharp edges and corners appear on the PNF surface and little gluing can be seen between each other. The average side length of the pyramids is ~48 nm and the average pore diameter is ~13 nm. When the substrate bias voltage is further increased to +100 V and +150 V, as shown in Fig. 1(c) and (d), both of the obtained films exhibit a triangular pyramids-like PNF structure more or less. The pyramids in Fig. 1(c) seem to glue to each other (like the case in Fig. (a)) while the pyramids in Fig. 1(d) show obvious edges and corners (like the case in Fig. 1(b)). Furthermore, the pyramids and pores in Fig. 1(c) and (d) are larger than those respectively in Fig. 1(b). For the average side length of pyramids: 71 nm in Fig. 1(c) and 69 nm in Fig. 1(d) vs. 48 nm in Fig. 1(b); for the average pore diameter: 21 nm in Fig. 1(c) and 20 nm in Fig. 1(d) vs. 13 nm in Fig. 1(b). When the substrate bias voltage further rises up to +200 V, as shown in Fig. 1(e), the film displays a typical "particle-void" porous structure with average diameters of particles and voids around 62 nm and 19 nm respectively. Based on the above observation, it can thus be concluded that the influence of substrate bias on the surface morphology is not linear. It implies that there may be two or more factors acting together or competing with each other.

### 3.2 Crystal structure

The GIXRD patterns in Fig. 2(a) show the crystal structure of $Cu_2O$ PNFs under different substrate bias voltage. Comparing with the standard card, it is found that the diffraction peaks in the patterns coincide well with those of $Cu_2O$ (card number: JCPDS-34-1354), and no other impurity peaks such as Cu and CuO are found in the patterns. It indicates that all the samples are pure $Cu_2O$ films, and the bias voltage does not change the chemical composition of the products. It can be further seen from the patterns that all the $Cu_2O$ films show a strong (111) preferred orientation, while the intensity of the diffraction peaks in other crystal orientations such as (110), (200), (220) and (311) are very weak and thus can almost be ignored. Since the out of plane GIXRD method was used in the experiment, it measured the crystal plane parallel to the surface of the film sample. This indicates that the orientation of $Cu_2O$ grains is almost the same in the growth direction of the films, all of which are (111) direction. In particular, we further study the relationships of the diffraction peak intensity and the grain size of $Cu_2O$ (111) with the positive bias voltage of the substrate, as shown in Fig. 2(b). It is found that, with the increase of substrate bias voltage, the diffraction peak intensity of $Cu_2O$ (111) decreases first and then increases, and reaches the minimum value at $V_s$ =+50 V. On the contrary, the calculation result by Scherrer Formula shows that the grain size of $Cu_2O$ (111) increases first and then decreases with the increase of substrate bias voltage, and reaches the maximum value at $V_s$ =+50 V. This complementary relationship may be because the number of grains in the film is inversely proportional to the size of grains. It is thus concluded that the influences of substrate bias on the diffraction peak intensity and the grain size are both not linear. Similar to the case of surface morphology, it further indicates that there may be two or more factors acting together or competing with each other.

### 3.3 Optical property

Fig. 3 gives the $(Ahv)^2 \sim hv$ curves of the $Cu_2O$ PNFs prepared under different substrate bias voltage, where $A$ is the absorbance of thin films and $hv$ is the energy of incident photons. Then the tangent line of the linear section in the $(Ahv)^2 \sim hv$ curve is made, and the extension line is given to the $hv$ axis. In doing so, their intersection point is the corresponding bandwidth $E_g$. It can be seen from the figure that when the substrate bias is $V_s$ =0 V, +100 V and +200 V, the values of the forbidden bandwidth are all 2.0 eV, while when the substrate bias is $V_s$ =+50 V and +150 V, the forbidden bandwidth values are 2.2 eV and 2.35 eV respectively. It demonstrates that $Cu_2O$ PNFs with different bandwidth can be prepared by adjusting the substrate bias voltage. Further, the above bandwidth values deviate from that of $Cu_2O$ bulk material (2.17 eV): red shift at $V_s$ =0 V, +100 V and +200 V; blue shift at $V_s$ =+50 V and +150 V. It is expected that the surface morphology and the crystallinity may both affect the bandwidth of PNFs. This is because the bandwidth of PNFs is not only related to the band structure of materials, but also affected by some other factors such as quantum size or nanosize effect, doping and defects [28]. In this work, firstly for different morphology, the nanosize effect of building nanostructures in PNFs will be accordingly different; secondly for different crystallinity, the effect of defects in PNFs on the bandwidth is also different. Based on the above morphological and structural findings, it can be found that the morphology and the crystallinity both influence the bandwidth of $Cu_2O$ PNF but in a nonlinear way. Further, we predict that the unique triangular pyramid structure of $Cu_2O$ PNFs with obvious edges and corners and less gluing (see Fig. 1(b,d)) may contribute more to the blue shift of energy band gap. On the contrary, the other PNF structures without obvious edges and corners or with more gluing (see Fig. 1(a,c,e)) may mainly lead to the red shift of energy band gap. At this level, the above optical experiment results are understandable although the specific influence mechanism is still unclear.

### 3.4 Bias deposition mechanism

Through the above analysis, we find that the surface morphology, crystal structure and optical property of the $Cu_2O$ PNF change accordingly when different positive bias voltage is applied on the substrate. This means that a certain effect of substrate bias plays an important role during the bias deposition of $Cu_2O$ PNF. In the existing literature, the research on bias deposition is mainly focused on the thin films prepared under a negative substrate bias in an unbalanced MS system [29-31]. One of the most notable feature of the unbalanced MS system is that part of its magnetic field lines can extend to the location of the substrate. Under the electric field of negative substrate bias, the positive sputtering ions such as $Ar^+$ can fly energetically to the substrate and then bombard the growing film effectively. As a result, a dense and solid film without porous structure is often obtained [29-31]. By contrast, almost all the magnetic field lines in the balanced MS system are bound near the target and thus only a very few sputtering ions or electrons can fly to the far substrate and bombard the growing film. As illustrated in Fig. 4(a), when a bias voltage $V_s$ is applied on the substrate, some of the sputtering ions (such as $Ar^+$, for negative bias voltage case) or electrons ($e^-$, for positive bias voltage case) may pass through the magnetic field of the target and reach the substrate. We measured the electric current $I$ (mA) in the circuit at different positive and negative bias voltage $V_s$ (V), which is shown in Fig. 4(b). Meanwhile, the electric current $I$ can be expressed using the following equation according to its definition:

$$I = nqvS \tag{1}$$

where $n, v$ are the concentration and velocity of $Ar^+$ or $e^-$ respectively near the substrate, $q=1.6\times10^{-19}$ C, and $S$ is the basal area of the 3-inch Cu substrate holder.

As driven by the electric field of negative or positive substrate bias, some of the $Ar^+$ or $e^-$ would be accelerated respectively and travel through the working gas in the chamber. Meanwhile, the $Ar^+$ or $e^-$ would collide with the gas atoms and degrade the energy on their way to the substrate. According to Eqn. (8) in ref. [32], the average retained energy $E_{k1}$ of $Ar^+$ or $e^-$ before reaching the substrate can be similarly expressed by the following equation:

$$E_{k_1} = \frac{1}{2}mv^2 = (qV_s + E_{k_0})\exp(-\alpha d / \lambda) \tag{2}$$

where $\alpha$ is a constant ($\alpha=0.835$ for the case of Cu target and Ar sputtering gas [32]), $d=15$ cm, represents the distance between the target and the substrate, $\lambda$ is the mean free path of $Ar^+$ or $e^-$ (8.06 cm for $Ar^+$ and $9.32\times10^9$ m for $e^-$ under an Ar pressure of 0.10 Pa [33]), and $E_{k0}$ is the initial kinetic energy of $Ar^+$ or $e^-$ when escaping from the plasma region. As opposed to tens or even hundreds of the substrate bias, here the initial kinetic energy $E_{k0}$ is negligible for simplicity. Thus, the simplified expression of the average retained energy $E_{k1}$ can be given as in the following:

$$E_{k_1} = \frac{1}{2}mv^2 = qV_s \exp(-\alpha d / \lambda) \tag{3}$$

According to Eqn. (3), the average retained energy $E_{k1}$ of $Ar^+$ ($E_{k1,Ar}$) or $e^-$ ($E_{k1,e}$) under different substrate bias is shown in Table 1. Combining Eqn. (1) and (3), we can further get the expression of the concentration of $Ar^+$ or $e^-$ close to the substrate:

$$n = \left[\frac{m}{2qV_s \exp(-\alpha d / \lambda)}\right]^{1/2} \cdot \frac{I}{qS} \tag{4}$$

According to Eqn. (4), we can calculate the concentration values of $Ar^+$ ($n_{Ar}$) and $e^-$ ($n_e$) close to the substrate at different substrate bias, which are shown in Table 1. It can be found that the order of magnitude of the concentrations are $10^{13}$ and $10^{15}$ m$^{-3}$ for $e^-$ and $Ar^+$ respectively. On the other hand, the concentration of Ar atoms in the working environment is $2.5\times10^{19}$ m$^{-3}$ by a simple calculation. In general, the ionization rate of sputtering gas is about 10% – 20%, which means that the concentration of $e^-$ or $Ar^+$ is $\sim10^{18}$ m$^{-3}$ in plasma. Obviously, either a positive bias or a negative bias, even as high as 200 V, the concentration of $e^-$ or $Ar^+$ close to the substrate is far less than that in plasma. It indicates that in the balanced MS system only a very small percentage of e⁻ or $Ar^+$ (~0.001% for e⁻ and ~0.1% for $Ar^+$) can escape from the localized magnetic field of the target and finally reach the substrate. Furthermore, the concentration of Cu atoms (will be further oxidized into $Cu_2O$ subsequently) or $Cu_2O$ molecules in the growing $Cu_2O$ film is estimated to be ~$10^{28}$ m$^{-3}$, which is much larger than that of the incident e⁻ ($10^{15}$ : 1) and $Ar^+$ ($10^{13}$ : 1). In this calculation, we assumed that the density of the growing $Cu/Cu_2O$ film is the same as that of bulk Cu and $Cu_2O$ respectively for simplicity. It demonstrates that only a very small portion of Cu atoms or $Cu_2O$ molecules in the growing film of $Cu_2O$ is bombarded by the e⁻ or $Ar^+$ on the biased substrate.

When the incident e⁻ or $Ar^+$ bombards the Cu atoms or $Cu_2O$ molecules on the substrate, the transfer of energy will be carried out. In this process, we assume that both energy and momentum are conserved for simplicity. Therefore, the transferred energy ($E_{k_2}$) to the Cu atoms or $Cu_2O$ molecules can be expressed by the following equation:

$$E_{k_2} = \frac{4m_1m_2}{(m_1+m_2)^2} E_{k_1} \tag{5}$$

where $m_1$ is the mass of an $Ar^+$ or an e⁻, and $m_2$ is the mass of a Cu atom or a $Cu_2O$ molecule. After a simple calculation, we can obtain the transferred energy to the Cu atoms ($E_{k2, Cu}$) or $Cu_2O$ molecules ($E_{k2, Cu2O}$), as listed in Table 1. When the positive substrate bias is applied, it is found that only a little energy in the range of (0.08 − 0.69)×$10^{-3}$ eV is transferred to the Cu atom or $Cu_2O$ molecule. Also considering the low concentration of incident e⁻, it can thus be concluded that the little bombardment of e⁻ will make not big enough and significant change to the deposited film. In contrast, when the negative substrate bias is applied, the energy in the range of 7.2 − 40 eV is transferred to the Cu atom or $Cu_2O$ molecule. It is expected that such a big energy transfer would cause the (self)-diffusion or even re-sputtering of the deposited Cu atoms or $Cu_2O$ molecules on the substrate. Nevertheless, due to the low concentration of incident $Ar^+$, it is also expected that a poor bombardment of $Ar^+$ will work during the film deposition. In contrast to the negative substrate bias, however, it is obviously a better choice to apply the positive substrate bias for the formation of PNFs due to the less bombardment effect and the resulting less opportunity to form dense films.

In the above, we have excluded the impact of bombardment of $Ar^+$ and e⁻ on the bias deposition of thin films, especially the bombardment of e⁻ during the positive bias deposition, in the balanced MS system. In the following, we propose a new model instead for the present positive bias deposition of $Cu_2O$ PNFs. As schematically illustrated in Fig. 5(a), since the film surface is uneven with tips or particles at the nanoscale, according to the tip charging effect, the charges induced by Coulomb force will gather preferentially at the tips or on the surface of particles. Thus, the electric field near the film surface can be regarded as a superposition of electric fields of the massive point charges. Under the electric field of point charges, the material atoms or molecules which fly to the substrate will be subject to the forces from two aspects. On one hand, the atoms or molecules of the material is polarized so that the centers of positive charge and negative charge appear to be separated (see Fig. 5(a)). Since the electric field of point charge is inhomogeneous (increases as the distance decreases), the polarized Cu atoms or $Cu_2O$ molecules will be attracted by the nearest point charges especially when they get close to the substrate (Force I). On the other hand, because of the repulsion between the same kind of charges, the atoms or molecules deposited at the tips and charged with the same charges will be excluded and bound by the electric field of the surrounding homologous charges (Force II). For the former, the point charge field provides an additional kinetic energy to the incoming atoms or molecules, which will promote the subsequent diffusion and migration on the surface of the substrate. For the later, the binding of the electric fields of the surrounding point charges will hinder the diffusion and migration of the material atoms or molecules between different tips or on the substrate surface. It must be pointed out that the tip charging effect exists in the entire deposition process as long as the positive bias voltage is applied. As a consequence, it universally induces a columnar growth of the film along the normal direction of the substrate (as evidenced by the FESEM image in Fig. 5(b), take the substrate bias of +50 V for example) and a significant $Cu_2O$ (111)-preferred orientation (see Fig. 2(a)) under the drive of reducing the system energy. However, the contradictory effects of Force I and II will exist together in each bias deposition process of thin films. The different contribution of these two effects at different substrate bias voltage will influence or even determine the final morphology and structure of the films. (1) When $V_s$ =0 V, there is no effect of substrate bias. The as-deposited Cu atoms or $Cu_2O$ molecules have no sufficient energy to ensure the effective diffusion and migration on the substrate, and also can not selectively deposit at the tip of the film surface. As a result, the surface of $Cu_2O$ film presents a randomly-oriented bicontinuous "ligament-channel" PNF structure. (2) When $V_s$ =+50 V and +150 V, Force II plays the major role. Under the constraint of the electric field of one's own locating tip, the Cu atoms or $Cu_2O$ molecules deposit and grow in a columnar form. Meanwhile, the adsorption of Cu atoms or $Cu_2O$ molecules is also affected by the neighboring tip fields, which likely leads to a lateral face-selective adsorption, thereby causing an anisotropic crystal growth of triangular pyramids with little adhesion to each other. (3) When $V_s$ =+100 V and +200 V, the contribution of Force I is greater than that of Force II. The atoms or molecules deposited at the tips will further diffuse along the surface of the film to a certain extent, so that the surface of the film is more glued and compacted.

In the balanced MS, we believe that the deposition mechanism for negative substrate bias is similar to that for positive substrate bias. However, there are still some differences, big or small, besides the above mentioned bombardment effect of $e^-$ and $Ar^+$. As we know, the mobility of electrons and holes in a semiconductor film is different. Consequently, they differ on the response speed under the positive and negative bias and the feedback effect of the tip charging. In general, the mobility of electrons is higher than that of holes. Thus, we infer that the response speed of electrons is faster than that of holes, and it is easier to gather electrons at the tips of the film, and thus the negative bias effect will be more obvious. Similarly, the carriers in the metal film are electrons, and the electrons can be more easily aggregated on the tips under the negative bias. In contrast, the metal cations are much more passive at the tips of the film surface under the positive bias. Based on this, we think that the effects of positive and negative bias on the tip charging effect or the bias deposition effect of the film will be different although a quantitative analysis is missing.

## 4. Conclusions

In this work, we successfully fabricated $Cu_2O$ PNFs on positively-biased substrates in a balanced MS system. It was found that the morphology, structure and optical property of the as-deposited products were non-linearly dependent on the applied positive substrate bias. In particular, when the substrate was biased at +50 V and +150 V, both of the obtained $Cu_2O$ PNFs exhibited a triangular pyramids-like structure with obvious edges and corners and little gluing, a preferred orientation of (111) and a blue shift of energy band gap at 2.35 eV. Quantitative calculation results indicated that the traditional bombardment effects of electrons and sputtering ions were both negligible during the bias deposition in the balanced MS system. Based on the experiment results and the related theoretical analysis, a new model of tip charging effect was further proposed to account for the flexible formation of PNFs by the balanced bias sputtering deposition.

## Acknowledgements

This work was financially supported by the NSFC project under grant no. 11574255 and the Students' Extracurricular Innovation and Entrepreneurship Fundation of Changzhou University under grants no. 2017-07-C-39 and 2018-07-C-60.

## References

1 Zhang JT, Liu JF, Peng Q, et al. Nearly monodisperse $Cu_2O$ and CuO nanospheres: preparation and applications for sensitive gas sensors. Chem Mater, 2006, 18: 867-871

2 Zhang H, Zhu Q, Zhang Y, et al. One-pot synthesis and hierarchical assembly of hollow $Cu_2O$ microspheres with nanocrystals-composed porous multishell and their gas-sensing properties. Adv Funct Mater, 2015, 17: 2766-2771

3 Han K, Tao M. Electrochemically deposited p-n homojunction cuprous oxide solar cells. Sol Energy Mater Sol Cells, 2009, 93: 153-157

4 Mittiga A, Salza E, Sarto F, et al. Heterojunction solar cell with 2% efficiency based on a $Cu_2O$ substrate. Appl Phys Lett, 2006, 88: 163502

5 Zheng Z, Huang B, Wang Z, et al. Crystal faces of $Cu_2O$ and their stabilities in photocatalytic reactions. J Phys Chem C, 2009, 113: 14448-14453

6 Yu H, Yu J, Mann S, et al. Template-free hydrothermal synthesis of $CuO/Cu_2O$ composite hollow microspheres. Chem Mater, 2007, 19: 4327-4334

7 Zhang Y, Deng B, Zhang T, et al. Shape effects of $Cu_2O$ polyhedral microcrystals on photocatalytic activity. J Phys Chem C, 2010, 114: 5073-5079

8 Zhu XF. Evidence of antisymmetry relation between a nanocavity and a nanoparticle: a novel nanosize effect. J Phys: Condens Matter, 2003, 15: L253-L261

9 Zhu XF, Wang ZG. Nanoinstabilities as revealed by shrinkage of nanocavities in silicon during irradiation. Int J Nanotechnol, 2006, 3, 492-516

10 Wang HH, Jiang MP, Su JB, et al. Fabrication of porous CuO nanoplate-films by oxidation-assisted dealloying method. Surf Coat Technol, 2014, 249: 19-23

11 Su JB, Jiang MP, Wang HH, et al. Microstructure-dependent oxidation-assisted dealloying of $Cu_{0.7}Al_{0.3}$ thin films. Russ J Electrochem, 2015, 51: 937-943

12 Figueiredo V, Elangovan E, Gonçalves G, et al. Effect of post-annealing on the properties of copper oxide thin films obtained from the oxidation of evaporated metallic copper. Appl Surf Sci, 2008, 254: 3949-3954

13 Ebisuzaki Y. Preparation of monocrystalline cuprous oxide. J Appl Phys, 1961, 32: 2027-2028

14 Dequan L, Zhibo Y, Peng W, et al. Preparation of 3D nanoporous copper-supported cuprous oxide for high-performance lithium ion battery anodes. Nanoscale, 2013, 5: 1917-1921

15 Wang L, Tao M. Fabrication and characterization of p-n homojunctions in cuprous oxide by electrochemical deposition. Electrochem Solid-State Lett, 2007, 10: 248-250

16 Ji J, Cooper WC. Electrochemical preparation of cuprous oxide powder: Part I. Basic electrochemistry. J Appl Electrochem, 1990, 20: 818-825

17 Li J, Shi Y, Cai Q. Patterning of nanostructured cuprous oxide by surfactant-assisted electrochemical deposition. Cryst Growth Des, 2008, 8: 2652-2659

18 Kamimura K, Sano H, Abe K, et al. Preparation of cuprous oxide ($Cu_2O$) thin films by reactive DC magnetron sputtering. Ieice T Electron, 2004, 87: 193-196

19 Hien VX, You JL, Jo KM, et al. H2S-sensing properties of $Cu_2O$ submicron-sized rods and trees synthesized by radio-frequency magnetron sputtering. Sensor Actuat B-Chem, 2014, 202: 330-338

20 Ishizuka S, Maruyama T, Akimoto K. Thin-film deposition of $Cu_2O$ by reactive radio-frequency magnetron sputtering. Jpn J Appl Phys, 2000, 39: 786-788

21 Lee YS, Winkler MT, Siah SC, et al. Hall mobility of cuprous oxide thin films deposited by reactive direct-current magnetron sputtering. Appl Phys Lett, 2011, 98: 192115

22 Masuda H, Fukuda K. Ordered metal nanohole arrays made by a two-step replication of honeycomb structures of anodic alumina. Science, 1995, 268: 1466-1468

23 Pal B, Sharon M. Enhanced photocatalytic activity of highly porous ZnO thin films prepared by sol-gel process. Mater chem phys, 2002, 76: 82-87

24 Erlebacher J, Aziz MJ, Karma A, et al. Evolution of nanoporosity in dealloying. Nature, 2001, 410: 450-453

25 Yang C, Jiang B, Liu Z, et al. Structure and properties of Ti films deposited by dc magnetron sputtering, pulsed dc magnetron sputtering and cathodic arc evaporation. Surf Coat Tech, 2016, 304: 51-56

26 Masłyk M, Borysiewicz MA, Wzorek M, et al. Influence of absolute argon and oxygen flow values at a constant ratio on the growth of Zn/ZnO nanostructures obtained by DC reactive magnetron

sputtering. Appl Surf Sci, 2016, 389: 287-293

27 Agnarsson B, Magnus F, Tryggvason TK, et al. Rutile $TiO_2$ thin films grown by reactive high power impulse magnetron sputtering. Thin Solid Films, 2013, 545: 445-450

28 Su JB, Zhang JH, Liu Y, et al. Parameter-dependent oxidation of physically sputtered Cu and the related fabrication of Cu-based semiconductor films with metallic resistivity. Sci China Mater, 2016, 59: 144-150

29 Thièry F, Pauleau Y, Ortega L. Effects of the substrate bias voltage on the physical characteristics of copper films deposited by microwave plasma-assisted sputtering technique. J Vac Sci Technol A, 2004, 22: 30-35

30 Seo SC, Ingram DC, Richardson HH. Effects of substrate bias on the properties of diamondlike carbon films deposited using unbalanced magnetron sputtering. J Vac Sci Technol A, 1995, 13: 2856-2862

31 Losbichler P, Mitterer C. Non-reactively sputtered TiN and $TiB_2$ films: influence of activation energy on film growth. Surf Coat Technol, 1997, 97: 567-573

32 Ekpe SD, Dew SK. Theoretical and experimental determination of the energy flux during magnetron sputter deposition onto an unbiased substrate. J Vac Sci Technol A, 2003, 21: 476-83

33 Westwood WD. Calculation of deposition rates in diode sputtering systems. J Vac Sci Technol, 1978, 15: 1-9

**Figures**

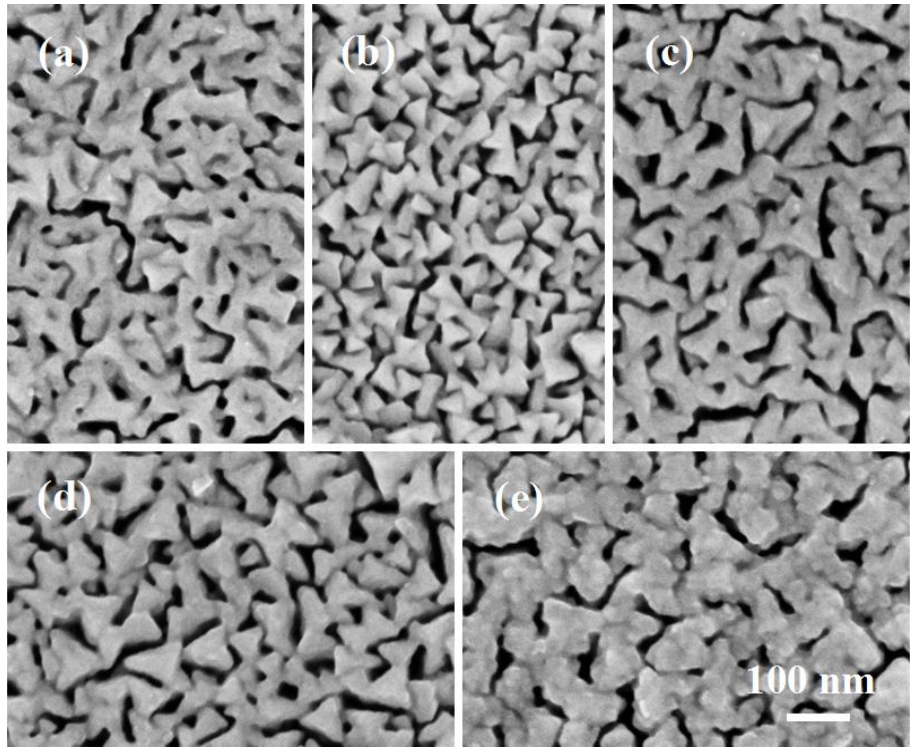

**Fig. 1** FESEM images showing the Cu$_2$O PNFs deposited at different positive substrate bias voltage: (a) 0 V; (b) +50 V; (c) +100 V; (d) +150 V; (e) +200 V.

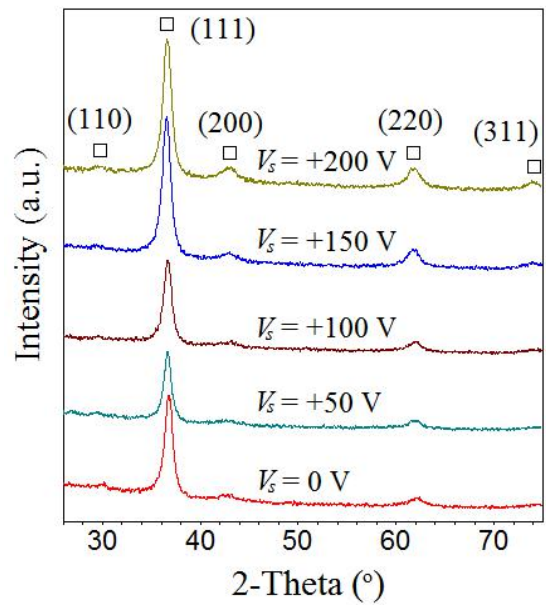

(a)

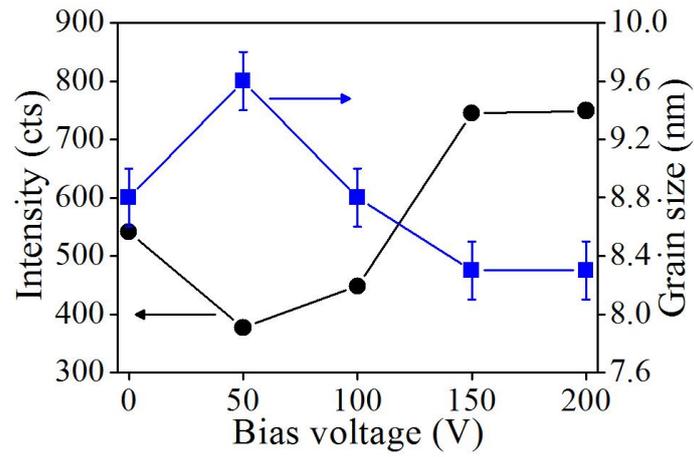

(b)

**Fig. 2** (a) GIXRD patterns of the Cu$_2$O PNFs obtained at different positive substrate bias voltage; (b) diffraction peak intensity and grain size of Cu$_2$O (111) against the positive substrate bias voltage.

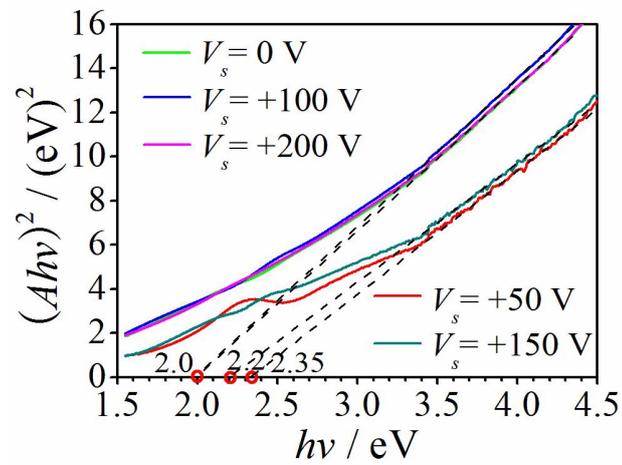

**Fig. 3** ($Ah\upsilon$)$^2$-$h\upsilon$ curves of the Cu$_2$O PNFs deposited at different positive substrate bias voltage.

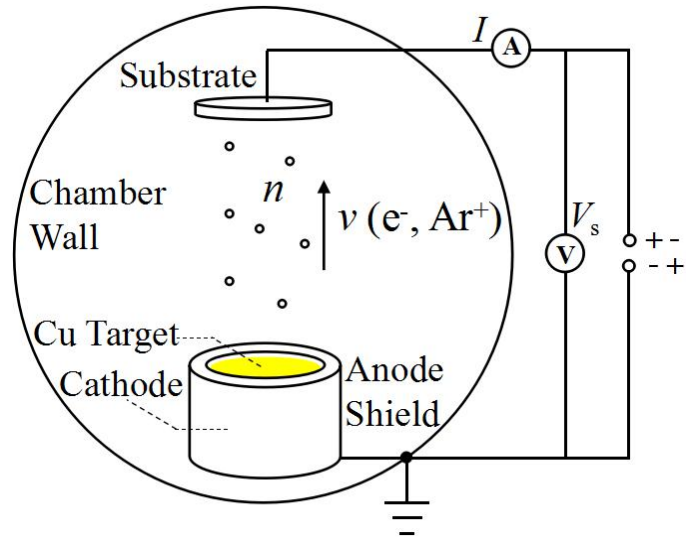

(a)

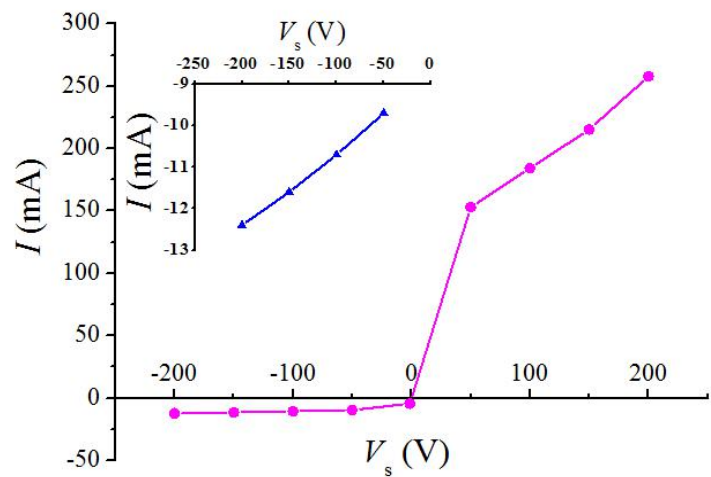

(b)

**Fig. 4** (a) Schematic diagram showing the installation for bias sputtering deposition of $Cu_2O$ PNFs; (b) the relationship between the applied substrate bias voltage $V_s$ (V) and the measured current $I$ (mA) in the circuit. The inset in figure (b) is the enlarged graph of the negative bias section (from -50 V to -200 V).

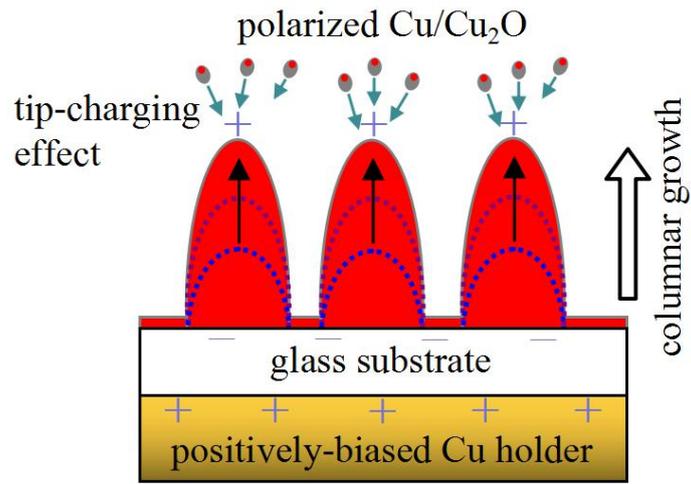

(a)

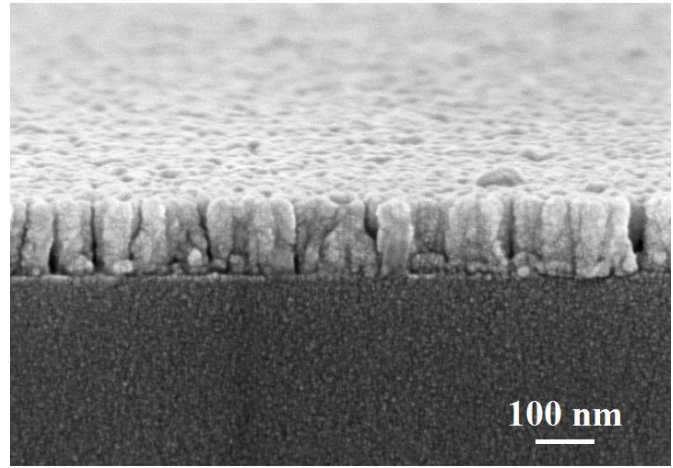

(b)

**Fig. 5** (a) Schematic illustration showing the columnar growth process of $Cu_2O$ PNFs by positive bias deposition; (b) cross-sectional view of the $Cu_2O$ PNFs obtained at substrate bias voltage of +50 V.

**Tables**

**Table 1** The concentration ($n_e$, $n_{Ar}$) and kinetic energy ($E_{k1,e}$, $E_{k1,Ar}$) of e⁻ and Ar⁺ close to the substrate and the transferred energy ($E_{k2,Cu}$, $E_{k2,Cu2O}$) to the Cu atoms and $Cu_2O$ molecules.

| $V_s$ (V) | -50 | -100 | -150 | -200 | +50 | +100 | +150 | +200 |
|---|---|---|---|---|---|---|---|---|
| $n_e$ (×10¹³ m⁻³) | / | / | / | / | 4.998 | 4.257 | 4.061 | 4.211 |
| $n_{Ar}$ (×10¹⁵ m⁻³) | 1.860 | 1.451 | 1.285 | 1.189 | / | / | / | / |
| $E_{k1,e}$ (eV) | / | / | / | / | 50 | 100 | 150 | 200 |
| $E_{k1,Ar}$ (eV) | 10.6 | 21.2 | 31.7 | 42.3 | / | / | / | / |
| $E_{k2,Cu}$ (eV) | 10.0 | 20.1 | 30.0 | 40.0 | 0.17×10⁻³ | 0.34×10⁻³ | 0.52×10⁻³ | 0.69×10⁻³ |
| $E_{k2,Cu2O}$ (eV) | 7.2 | 14.4 | 21.5 | 28.7 | 0.08×10⁻³ | 0.15×10⁻³ | 0.23×10⁻³ | 0.30×10⁻³ |